\def\rn{\noindent\parshape 2 0truecm 8.8truecm 0.3truecm 8.5truecm}
\def\nn#1 #2{#1, #2.}				
\def\nnn#1 #2 #3{#1, #2. #3.}			
\def\nnnn#1 #2 #3 #4{#1, #2. #3. #4.}		
\def\dualand{, \&\hbox{ }}				
\def\multiand{, \&\hbox{ }}				
\def\rfprep#1;#2;#3 {{\par\rn#1 #2, #3\par}}

\def\rg#1;#2;#3;#4;#5;#6 {\par\rn#1 #2, {\it #3}, {\bf #4}, #5 (``#6'') \par}
\def\rf#1;#2;#3;#4;#5 {\par\rn#1 #2, {\it #3}, {\bf #4}, #5\par}
\def\rfbook#1;#2;#3;#4;#5 {{\frenchspacing\par\rn#1 #2, {\it #3} (#4: #5)\par}}
\def\rfproc#1;#2;#3;#4;#5;#6 {{\frenchspacing\par\rn#1 #2, in {\it #3}, ed. #4 (#5: #6)\par}}

\def\K{{\rm K}}

\def\muK{{\rm \mu K}}
\def\MJy{{\rm MJy}}

\def\sr{{\rm sr}}
\def\MJysr{\MJy/\sr}

\def\GHz{{\rm GHz}}

\def\expec#1{\langle#1\rangle}

\def\etal{{\frenchspacing\it et al.}}
\def\ie{{\frenchspacing\it i.e.}}
\def\eg{{\frenchspacing\it e.g.}}

\def\rms{rms}

\def\beq#1{\begin{equation}\label{#1}}
\def\eeq{\end{equation}}
\def\beqa#1{\begin{eqnarray}\label{#1}}
\def\eeqa{\end{eqnarray}}
\def\eq#1{equation~(\ref{#1})}
\def\Eq#1{Equation~(\ref{#1})}

\def\ns{\vskip-0.2truecm}

\def\spose#1{\hbox to 0pt{#1\hss}}
\def\simlt{\mathrel{\spose{\lower 3pt\hbox{$\mathchar"218$}}
     \raise 2.0pt\hbox{$\mathchar"13C$}}}
\def\simgt{\mathrel{\spose{\lower 3pt\hbox{$\mathchar"218$}}
     \raise 2.0pt\hbox{$\mathchar"13E$}}}
\def\simpropto{\mathrel{\spose{\lower 3pt\hbox{$\mathchar"218$}}
     \raise 2.0pt\hbox{$\propto$}}}

\def\ed{\end{document}}

\def\FWHM{{\rm FWHM}}
\def\alm{a_{\l m}}
\def\Ylm{Y_{\l m}}
\def\Cltot{C^{tot}_\l}
\def\Clnoise{C^{noise}_\l}
\def\Cps{C^{ps}}
\def\sigps{\sigma_{ps}}
\def\sumlm{\sum_{\l m}}
\def\Tcmb{T_{cmb}}
\def\lfac{\left({2\l+1\over 4\pi}\right)}
\def\ith{i^{th}}
\def\l{\ell}
\def\rh{\widehat{\bf r}}
\def\x{\eta}

\documentstyle[emulateapj,epsf,rotate]{article}
\begin{document}


\journalid{337}{15 January 1989}
\articleid{11}{14}

\submitted{Submitted to ApJL February 11, 1998; accepted April 14;
published June 8}

\title{REMOVING POINT SOURCES FROM CMB MAPS}

\author{
Max Tegmark
\footnote{Institute for Advanced Study, Princeton, 
NJ 08540; max@ias.edu}$^,$\footnote{Hubble Fellow}
and
Ang\'elica de Oliveira-Costa\footnote{
Princeton University, Department of Physics, Princeton, NJ 08544; angelica@ias.edu}$^{,1}$
}



\begin{abstract}
For high-precision cosmic microwave background 
(CMB) experiments, contamination from extragalactic 
point sources is a major concern. It is therefore useful to be able
to detect and discard point source contaminated pixels using the 
map itself. 
We show that the sensitivity with which this can be done
can often be greatly improved (by factors between 2.5 and 18 for 
the upcoming Planck mission) by a customized 
hi-pass filtering that suppresses fluctuations due to 
CMB and diffuse galactic foregrounds. 
This means that point source contamination will not severely degrade
the cleanest Planck channels unless current source count estimates are off
by an order of magnitude.
A catalog of around 40,000 far infra-red sources at 857 GHz may be a 
useful by-product of Planck. 
\end{abstract}


\makeatletter
\global\@specialpagefalse
\def\@oddfoot{
\ifnum\c@page>1
  \reset@font\rm\hfill \thepage\hfill
\fi
\ifnum\c@page=1
{\sl
Available in color from
h t t p://www.sns.ias.edu/$\tilde{~}$max/pointsources.html}
\hfill\\
\fi
} \let\@evenfoot\@oddfoot
\makeatother


\section{INTRODUCTION}

If future CMB experiments are to produce high-precision 
measurements of cosmological parameters 
(Jungman {\etal} 1996; Bond {\etal} 1997; Zaldarriaga {\etal} 1997),
they must remove foreground contamination from
galactic dust, synchrotron and free-free emission as well as
extragalactic point sources with comparable accuracy
(see {\eg} Brandt {\etal} 1994; Tegmark \& Efstathiou 1996 -- hereafter TE96;
Bersanelli {\etal} 1996).
Fortunately, foregrounds differ from CMB fluctuations in several 
ways, all of which can be used as weapons against them,
in combination:
\renewcommand{\theenumi}{\roman{enumi}}
\begin{enumerate}
\itemsep0cm
\item Their non-Gaussian behavior can be used to discard 
severely contaminated regions 
(\eg, bright point sources, the Galactic plane).

\item Their frequency dependence can be used to subtract them out by 
taking linear combinations of maps at different frequencies.

\item Their power spectra can be compared with the one measured,
to fit out remaining foregrounds.

\end{enumerate}

Extragalactic point sources are one of the most 
menacing foregrounds, for several reasons:
\begin{itemize}
\itemsep0cm
\item Their spectral index varies much more than for other foregrounds.

\item Many sources exhibit substantial time variability.

\item Their abundance is very poorly known over much of the CMB frequency range.
\end{itemize}
The variation of their frequency dependence in a random way across the sky 
(from source to source) 
substantially degrades the effectiveness of weapon (ii) (Tegmark 1998), 
and time-variability of both flux and spectral shape
(\eg, Gutierrez de la Cruz {\etal} 1998) further 
complicates subtraction attempts.
As a complement to (ii), 
it is therefore important to make as much use as possible of 1.
This is the purpose of the present {\it Letter}.

If point sources give the only non-Gaussian contribution to a map, 
then a useful way to implement 1 is to discard all pixels
whose temperature lies more than $\nu$ standard deviations $\sigma$ 
above the mean.
Chosing $\nu=5$ ensures that this will falsely reject only 
a negligible fraction of order $3\times 10^{-7}$ of all uncontaminated
pixels. 
The accuracy with which this method will be able to clean 
the CMB maps from the upcoming Planck satellite (Bersanelli et al 1996)
has recently been estimated by Gawiser \& Smoot (1997), 
Toffolatti et al (1998, hereafter To98) and Guiderdoni {\etal} (1998).
Similar estimates have been made for the MAP experiment
(Refregier {\etal} 1998).
For such high signal-to-noise experiments, the {\rms} pixel
fluctuations $\sigma$ are not dominated by detector noise 
but by CMB (or, in some channels, by galactic foregrounds).
We will show that much of these fluctuations (which hamper
our ability to detect point sources) can be removed by  
an appropriately chosen band-pass filter, and that consequently, 
the threshold $\nu\sigma$ at which point sources can be removed 
can be substantially lowered.
We derive our method in \S2, apply it to Planck in \S3 and
discuss our results in \S4.

\section{METHOD}

If there are point sources with fluxes $S_i$ at sky positions given
by unit vectors $\rh_i$, then the resulting 
sky temperature $x$ is
\beq{xEq}
x(\rh) = c\sum_i S_i\delta(\rh_i,\rh) + \sumlm\alm\Ylm(\rh),
\eeq
where $\delta$ is a Dirac delta function and 
the spherical harmonic coefficients $\alm$ contain the combined 
contributions from 
CMB, galactic foregrounds and detector noise.
Here $c$ is the conversion factor between surface brightness and temperature, 
given by (see, \eg, equation (3) in TE96)
\beq{cEq}
c = c_* {(2\sinh{\x\over 2})^2\over\x^4},
\quad
c_* \equiv
{1\over 2k} \left({hc\over k\Tcmb}\right)^2
\approx
{10\,{\rm mK}\over\MJysr}
\eeq
where $\x\equiv h\nu/k\Tcmb\approx \nu/56.8\,\GHz$.
The observed map is $x(\rh)$ convolved with the beam function $B$.
To maximize our sensitivity to the point sources, we wish to 
convolve it with an additional function $W$,
giving a filtered map
$y(\rh)\equiv(W\star B\star x)(\rh)$ ($\star$ denotes convolution).
\Eq{xEq} gives
\beq{yEq}
y(\rh)
=c\sum_i S_i(W\star B)(\rh_i\cdot\rh) + \sumlm W_\l B_\l\alm\Ylm(\rh).
\eeq
Here we have taken both the filter $W$ and the beam $B$ to be 
spherically symmetric, so that they are given by the coefficients
$W_\l$ and $B_\l$ in a Legendre polynomial expansion:
$W(\cos\theta)=\sum_\l\lfac W_\l P_\l(\cos\theta)$ and 
$B(\cos\theta)=\sum_\l\lfac B_\l P_\l(\cos\theta)$.
\Eq{yEq} tells us that if the contribution from
overlap of nearby sources is negligible
(we will see that this is a good approximation for Planck
at the attainable flux threshold), 
then the point source contribution in the direction of the
$\ith$ source is 
$y(\rh_i)=A S_i$, where the normalization constant 
$A\equiv (W\star B)(1)c$, since $\rh_i\cdot\rh_i=1$.
This constant can be re-written as 
\beq{Aeq}
A = (W\star B)(1)c = c\sum_\l\lfac B_\l W_\l.
\eeq
In other words, \eq{yEq} tells us that the peak brightness of a source 
in the normalized map $y(\rh)/A$ gives its strength $S$ in
flux units (Janskys). This is true for any choice of 
our filter $W$, so we simply wish to choose $W$ so that it minimizes
the variance $\sigma^2$ in this map.
Equivalently, we want to maximize the signal-to-noise ratio,
which is the ratio of the peak signal of a source after filtering
to the {\rms} fluctuation level $\sigma$ 
in this region not due to the source.
These fluctuations come from the $\alm$ in \eq{yEq}, 
\ie, from CMB, pixel noise and galactic foregrounds, 
all which act as unwanted noise now that point source detection
is our objective.
Modeling these as Gaussian random fields as in TE96, we have 
$\expec{\alm^* a_{\l'm'}}=\delta_{\l\l'}\delta_{mm'}\Cltot$,
where $\Cltot$ is the sum of the power spectra of the
CMB, the noise and various Galactic foreground components.
Using this, \eq{yEq} and \eq{Aeq}, we find that the 
variance of our point source map $y(\rh)/A$ is 
\beq{sigmaEq}
\sigma^2\equiv V\left[y(\rh)\over A\right] = 
c^2{\sum_\l\lfac B_\l^2\Cltot W_\l^2\over\left[\sum_\l\lfac B_\l W_\l\right]^2}.
\eeq
It is easy to show that this variance is minimized if we choose the
filter to be
\beq{Weq}
W_\l\propto {1\over B_\l\Cltot},
\eeq
which gives
\beq{MinimalSigmaEq}
\sigma^2 = c^2\left[\sum_\l\lfac/\Cltot\right]^{-1}.
\eeq
In summary, the outlier removal method can eliminate
all point sources with flux $S>S_c\equiv\nu\sigma$, where $\sigma$
is given by \eq{MinimalSigmaEq}, and $\nu$ can be chosen depending 
on the desired confidence level of source detection.

We note that our assumption of symmetric beams is by no means necessary.
Indeed, our filtering method is readily generalized to the case
of arbitrary beam shape and arbitrary known profiles of 
(resolved) sources, \ie, galaxy clusters,
in a manner analogous to Appendix A of Haehnelt \& Tegmark (1996).

\section{APPLICATION TO PLANCK}

We will now illustrate our method with an application to the upcoming 
Planck mission, to see by what factor it reduces $\sigma$.

\ns\ns\ns
\subsection{Assumptions about foregrounds, noise and CMB}
 
For the Galactic foreground power spectra, we use the model 
$C_\l(\nu)\propto c^2 B(\nu)^2\l^{-3}$ (TE96; Bersanelli {\etal} 1996)
for dust, synchrotron and free-free emission,
where $B$ is the sky brightness measured in $\MJy/\sr$.
We model the frequency dependence as 
$B(\nu)\propto \nu^{-0.15}$ for free-free emission
and as $B(\nu)\propto \nu^{-0.9}$ for synchrotron emission
(see Platania {\etal} 1998 and references therein).
For dust, we assume $B(\nu)\propto\nu^{3+\beta}/(e^{h\nu/kT}-1)$,
with a dust temperature $T=20\K$ and an emissivity $\beta=1.5$
(Kogut {\etal} 1996, hereafter K96).

The combined DIRBE and IRAS dust maps
suggest a slightly shallower slope $\l^{-2.5}$ (Schlegel {\etal} 1998),
but a recent analysis of the DIRBE maps has shown no evidence
of a departure from an $\l^{-3}$ power law (Wright 1998) for 
$\l\simlt 300$, and we will see that only the behavior 
at low $\l$ matters for the present analysis. 

We normalize the power spectra based on the DIRBE-DMR cross-correlation 
analysis of K96, which gives
{\rms} fluctuations of $2.7\muK$ for dust and $7.1\muK$ for
free-free emission at 53 GHz on the COBE angular scale, in 
good agreement with the DIRBE cross-correlation results for
the Saskatoon map (de Oliveira-Costa {\etal} 1997)
and 19 GHz map (de Oliveira-Costa {\etal} 1998).
We normalize the synchrotron model to give $11\muK$ at 31 GHz, 
the K96 upper limit,
since cross-correlation between the 408 MHz and 19 GHz emission
indicates a value in this range
(de Oliveira-Costa {\etal} 1998).
For radio and infrared point sources, 
we use the source count model of To98.

The noise power spectrum is $\Clnoise=(\FWHM\sigma_n)^2/B_\l^2$
(Knox {\etal} 1995; TE96), where $\sigma_n$ is the {\rms} noise 
in a pixel of area $\FWHM^2$. 
We assume Gaussian beams of {\rms} width $\theta$, which corresponds to
$B_\l = e^{\theta^2\l(\l+1)/2}$, where $\theta=\FWHM/(8\ln 2)^{1/2}$.
The values of $\nu$, $\FWHM$ and $\sigma_n$ for the Planck channels 
are taken from {\it http://astro.estec.esa.nl/Planck/}. 

We compute the CMB power spectrum with CMBFAST 
(Seljak \& Zaldarriaga 1996) for three COBE-normalized 
models.
We use a standard cold dark matter model
(SCDM) with $n=1$, $h=0.5$, $h^2\Omega_b=0.015$
as well as two models that
agree better with observational data (Wang {\etal} 1998).
These are $\Lambda$CDM, a flat model with a cosmological constant
(as SCDM except that $\Omega_\Lambda=0.5$),
and OCDM, an open model 
(as SCDM except that $\Omega_\Lambda=0$, $\Omega=0.5$, $h=0.65$).

\ns\ns\ns
\subsection{Results}

The optimal filter $W$ for Planck channel 5 in the 
$\Lambda$CDM model is shown in real space
in Figure 1 and in the Fourier (multipole) domain in Figure 2.
The $\Lambda$CDM 
results for all channels are summarized in Table 1,
and it is seen that the sensitivity $\sigma$ is improved
by a factor ranging from 2.5 to 18.
This can be qualitatively understood from Figure 2, which 
shows that $W$ is essentially a band-pass filter, targeting
those multipoles where the combined fluctuations from CMB, foregrounds 
and noise are minimal. 
The table shows that the dominant sky signal
(CMB for channels 1-7, dust for channels 8-10)
is suppressed by even larger factors, 
at the cost of higher detector noise.
The filtering technique therefore helps the most
for high resolution experiments where the signal-to-noise level
per resolution element is substantially greater than unity.
This should be attainable for instance for several of the upcoming 
interferometer experiments; see Smoot (1997) for a recent review.
Conversely, the filtering helps only marginally (with gains 
below a factor of 2)
for the upcoming MAP satellite, for which we repeated our analysis
using the specifications from 
{\it http://map.gsfc.nasa.gov}. 
This is because the MAP detector noise is not too far below 
the fluctuation levels from CMB and foregrounds to start with, 
leaving less room than in the Planck case
for filtering to improve the situation.
The same conclusion is drawn in the MAP analysis of Refregier {\etal} (1998).

Table 1 shows that the number of sources removed
changes by an even greater factor than $\sigma$ does.
This is because the differential source counts
of To98 are quite steep. Channel 10
is especially notable: here Planck should be able to produce 
a catalog with around 40,000 sources, undoubtedly useful in its
own right, as compared to a measly 300
without filtering. Our filtering technique should
also be useful for constructing a catalog of SZ-clusters
from the 217 GHz map, as described by Aghanim {\etal} (1997).

The SCDM and OCDM models give results quite similar to
Table 1, with gain factors differing by less than $20\%$.

\vskip-1.7cm
\centerline{\rotate[l]{\vbox{\epsfxsize=7.7cm\epsfbox{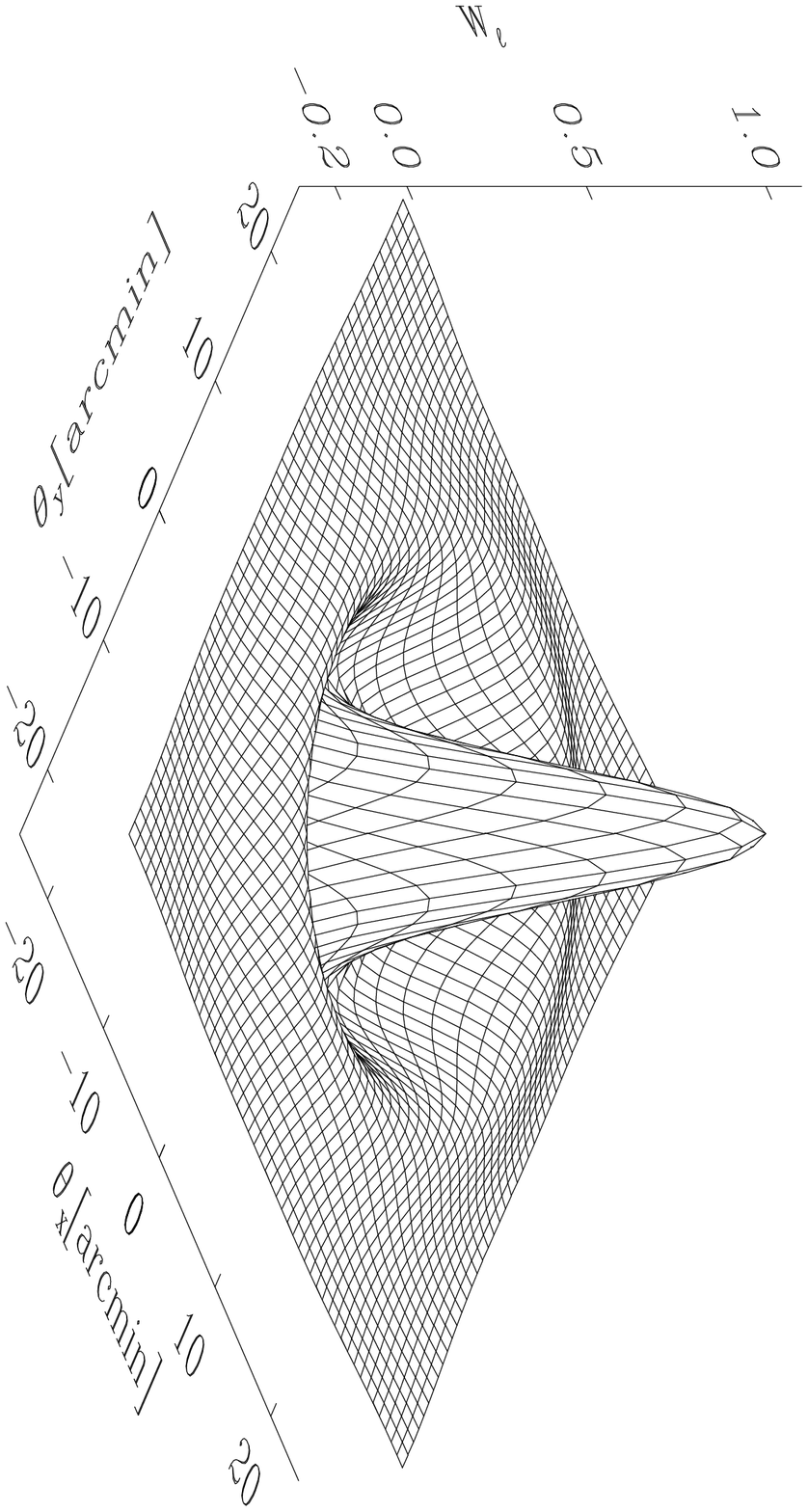}}}}
\vskip-0.5cm
{\footnotesize {\bf FIG. 1} 
--- The convolution filter $W$ is plotted in real space for 
channel 5 of Planck.
}

\section{DISCUSSION}

We have presented a method
for point source removal
that consists of the following steps:
Convolve the CMB map with a band-pass filter such as the one
in Figure 1.
Compute the resulting $\rms$ fluctuation level $\sigma$ 
and interpret all positive fluctuations exceeding $\nu\sigma$
as point sources. Revert to the original map and discard 
the contaminated pixels (within a beam size or two of each source).

The convolution kernel $W$ is typically well-localized on the sky,
which means that the convolution can be computed 
directly with a reasonable effort.
Alternatively, the convolution can be performed by 
expanding the sky-map in spherical harmonics, 
multiplying the expansion coefficients $\alm$ by $W_\l$,
and transforming back.
If necessary, the high-pass filtering can be further accelerated 
by making a flat-sky approximation locally and performing the 
convolution using two-dimensional fast Fourier transforms.
All these cases involve performing sums rather than integrals,
since  real-world maps are discretized into pixels of
finite size. This discreteness should not substantially
degrade the foreground removal as long as the
map is properly oversampled: Figure 2 shows that the
filter has no power below the beam size, where $\Clnoise$ blows up,
and therefore will not vary appreciable from one pixel to the next.

We found that this simple method 
can enhance the detectability of point sources by a substantial factor,
which mainly depends on the available signal-to-noise in the 
unfiltered map. The detectability can of course be further improved by taking 
an appropriate linear combination of maps at different frequencies
(as in Tegmark 1998) 
before the filtering, to remove {\eg} the CMB
at that stage at the cost of 
a slight noise increase.

Our filtering procedure has an additional advantage.
Since the outlier removal scheme is more likely to
throw away pixels in CMB hot spots than in cold spots, 
it inevitably introduces a slight bias, 
correlating false positives and false negatives 
with the CMB. 
With a $5-\sigma$ threshold, 
a $3-\sigma$ point source will get removed if it resides in a 
$2-\sigma$ CMB hot spot, whereas it takes a $7-\sigma$ point source
to be detected in a corresponding CMB cold spot.
By eliminating most of the CMB before the point source removal 
step, our filtering scheme also eliminates most of this bias.

\vskip-1.0cm
\centerline{{\vbox{\epsfxsize=9.7cm\epsfbox{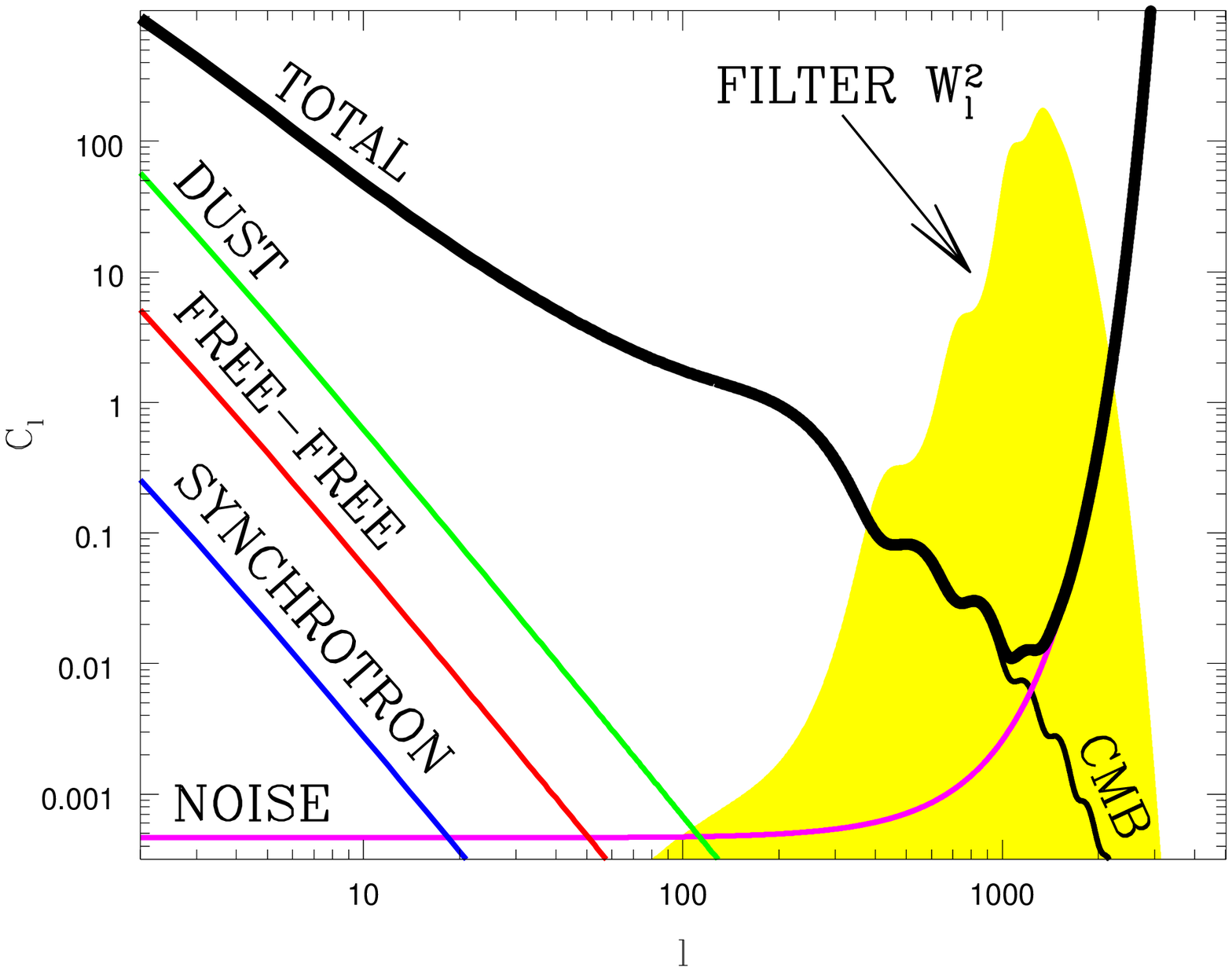}}}}
\vskip-1.1cm
{\footnotesize {\bf FIG. 2} 
--- Same as Figure 1, but in the Fourier
(multipole) domain. $W_\l^2$ is plotted
(shaded) together with the total power spectrum
$\Cltot$ (heavy line) and the various components 
that make it up. Everything is for Planck channel 5,
\ie, at 100 GHz.
}


\ns
\subsection{Robustness of the results}

How sensitive are our results to the various assumptions that we have made?
We found that the Galactic foreground model made only a minimal difference
for most Planck channels, where the contribution to $\Cltot$ 
(and hence to the confusion noise $\sigma$) 
is likely to be dominated by CMB and detector noise for all $\l$.
Likewise, we found that changing the cosmological model
had only a minor ($<20\%$) effect on $\sigma$.
MAP and Planck can of course use the model determined 
by their data, iteratively.
Instead, almost all the uncertainty comes from the assumed source
count model (To98), so let us compute this dependence explicitly.
If we can remove all sources brighter than a flux cut $S_c$, 
the point source power spectrum in the original 
(unfiltered) map is (TE96; Tegmark \& Villumsen 1997)
\beq{CpsEq}
{1\over c^2}\Cps = \int_0^{S_c} n'(S)S^2dS
\approx\left(\beta-1\over 3-\beta\right)n(S_c) S_c^2,
\eeq
independent of $\l$,
where $n(S)$ is the number of point sources per steradian 
with a flux exceeding $S$ and 
$\beta\equiv -d\ln n'/d\ln S$ is the logarithmic slope of
the differential source counts $n'\equiv -dn/dS$.
This neglects point source clustering, which 
is generally a good approximation (To98).
The {\rms} fluctuations $\sigps$ are given by the familiar
expression
\beq{sigpsEq}
\sigps^2= \Cps\sum_\l\lfac B_\l^2\approx{1\over 4\pi\theta^2}\Cps,
\eeq
where the sum can be accurately approximated by an integral.
Above we saw that outlier removal gave 
$S_c=\nu\sigma/cB(1)=2\pi\theta^2\nu\sigma/c$.
Substituting this and \eq{CpsEq} into 
\eq{sigpsEq}, we obtain the useful result 
\beq{sigpsApproxEq}
\sigps\approx\sqrt{\beta-1\over 3-\beta}N^{1/2}\nu\sigma,
\eeq
where $N\equiv\pi\theta^2 n(S_c)$ is the number of sources
removed per beam area.
Since relevant values for $\beta$ are typically in the range 1.5--2.5
(see references in Tegmark \& Villumsen 1997), 
the first term is of order unity.
Table 1 showed that the best 
attainable $\sigma$ was typically 3-5 times the {\rms}
noise $\sigma_n$.
Point sources have only a minor impact on a CMB 
experiment if $\sigps\ll\sigma_n$, since their power spectra
have the same shape as that for detector 
noise (apart from the noise blowup up the beam scale).
\Eq{sigpsApproxEq} therefore tells us that 
using the CMB map itself for point source removal is quite adequate
as long as $N\ll (4\times 5)^{-2}=0.002$. 
Conversely, if there are more sources per beam than this rule of 
thumb indicates, then an external point source template will
be needed to reduce the point source contribution to a
subdominant level.
This criterion thus
partitions CMB experiments into two classes: those for which 
internal cleaning suffices and those which need external point 
source data.

Since $\sigps\propto N^{1/2}$, a useful measure is the {\it safety margin},
defined as $M\equiv(\sigma_n/\sigps)^2$. This is the factor by which 
the number of point sources can be increased before they dominate
the noise {\rms}. In Table 1, it is seen to be of order 10
for channels 7 and 8, which means that the models of To98 would need to
be off by an order of magnitude for point sources to imperil 
the Planck mission. For those channels where $M\simlt 1$
($\sigps\simgt\sigma_n$), it will be desirable to use a multi-frequency subtraction 
scheme or external point source catalogs to further reduce $\sigps$.
Future radio source surveys at CMB frequencies will therefore be
extremely valuable to the CMB community.

\smallskip
The authors wish to thank Alexandre Refregier
for stimulating discussions, Luigi Toffolatti for kindly
providing source count model data from To98, and 
Uro\v s Seljak \& Matias Zaldarriaga for use
of their CMBFAST code.
Support for this work was provided by
NASA though grant NAG5-6034 and 
a Hubble Fellowship,
HF-01084.01-96A, awarded by STScI, which is operated by AURA, Inc. 
under NASA
contract NAS5-26555



\ns\ns\ns\ns

\nobreak


\def\freq{$\nu$}
\def\eff {$_{\rm eff}$}
\def\fwhm{\footnotesize FWHM}
\def\sign{$\sigma_{{\rm n}}$}
\def\sigg{$\sigma_{{\rm Gal}}$}
\def\sigp{$\sigma_{{\rm ps}}$}
\def\sigm{\footnotesize {\rm M}}
\def\sigcmb{$\sigma_{{\rm CMB}}$}
\def\slim{$S_{{\rm c}}$}
\def\nlim{$N(>5\sigma)$}
\def\mjy{mJy}
\def\A{\footnotesize A}
\def\B{\footnotesize B}

\ns\ns

{\footnotesize
\begin{center}
\begin{tabular}{cccccccccccccccccc}
\hline
\hline
\multicolumn{1}{c}{\freq\eff}&
\multicolumn{1}{c}{\fwhm}    &
\multicolumn{2}{c}{\sign}    &
\multicolumn{2}{c}{\sigg}    &
\multicolumn{2}{c}{\sigcmb}  &
\multicolumn{2}{c}{$\sigma$}    &
\multicolumn{1}{c}{Gain}     &
\multicolumn{2}{c}{\nlim}    &
\multicolumn{1}{c}{\sigp}    &
\multicolumn{1}{c}{\sigm}   \\

\multicolumn{1}{c}{(GHz)}  &
\multicolumn{1}{c}{(')}    &
\multicolumn{2}{c}{(\mjy)} &
\multicolumn{2}{c}{(\mjy)} &
\multicolumn{2}{c}{(\mjy)} &
\multicolumn{2}{c}{(\mjy)} & &
\multicolumn{2}{c}{(8 sr)} & 
\multicolumn{1}{c}{(\mjy)}\\

\cline{ 3-10}
\cline{12-15}

 & &
\multicolumn{1}{c}{\B} &
\multicolumn{1}{c}{\A} &
\multicolumn{1}{c}{\B} &
\multicolumn{1}{c}{\A} &
\multicolumn{1}{c}{\B} &
\multicolumn{1}{c}{\A} &
\multicolumn{1}{c}{\B} &
\multicolumn{1}{c}{\A} &  &
\multicolumn{1}{c}{\B} &
\multicolumn{1}{c}{\A} &
\multicolumn{1}{c}{\A} &
\multicolumn{1}{c}{\A}\\
\hline
 30&   33&  12&  59&   80&   2&  242&   66&  255&  89& 2.9& 178&   867&  46& 0.06 \\
 44&   23&  19&  71&   36&   1&  271&   77&  274& 105& 2.6& 156&   674&  33&  0.3 \\
 70&   14&  25&  78&   16& 0.5&  257&   73&  262& 106& 2.5& 165&   655&  19&  1.5 \\
100&   10&  27&  79&   19& 0.5&  247&   62&  252& 101& 2.5& 175&   697&  13&  3.8 \\
100& 10.6&  13&  51&   21& 0.4&  279&   48&  280&  70& 4.0& 147&  1157&  12&  1.1 \\
143&  7.4&  11&  39&   31& 0.6&  225&   31&  227&  51& 4.5& 131&  1220&   6&  3.0 \\
217&  4.9&  14&  36&   51&   1&  130&   20&  141&  41& 3.4& 206&  1288&   4& 14.8 \\
353&  4.5&  24&  50&  186&   7&   69&   20&  199&  54& 3.7& 205&  1712&   8&  9.0 \\
545&  4.5&  45&  87&  609&  32&   13&    6&  611&  93& 6.5& 209&  4696&  24&  3.6 \\
857&  4.5&  36&  82& 1735&  49&  0.3& 0.06& 1735&  95&  18& 293& 37612& 111&  0.1 \\
\hline
\end{tabular}
\end{center}
}

\nobreak
\vbox{
{\footnotesize 
{\bf Table~1} --- The rows show how the 10 Planck channels can be used
for point source detection before (B) and after (A) filtering.
The values of {\rms} confusion noise 
correspond to detector noise (\sign), 
Galactic foregrounds ({\sigg} is the combined contribution of 
dust, free-free and synchrotron emission),
CMB (\sigcmb), the quadrature sum of all of the above ($\sigma$)
and unremoved point sources (\sigp).
{\sigcmb} is computed from
the full power spectrum -- using 
\sigcmb$=10^{-5}T_{cmb}\approx 27\muK$ as in {\eg} To98 gives
values about 5 times too low. The gain is the factor by which filtering
reduces $\sigma$. {\nlim}
is the number of detected sources in 8 sr (having $S>S_c$)
and $M$ the factor by which the To98 source counts would have to be 
increased to give {\sigp=\sign}.

}
}

\goodbreak

\ed
\bye